\documentclass[reprint, amsmath, amssymb, amsfonts, aps]{revtex4-2}

\usepackage[english]{babel}
\usepackage{graphicx}
\usepackage{xcolor}
\usepackage[uncertainty-mode=separate]{siunitx}
\usepackage{textcomp}

\newcommand{\figref}[1]{Figure~\ref{#1}}
\newcommand{\id}[1]{\ensuremath{\mathrm{d}#1}}

\begin{document} 

\title{Laser-FLASH: radiobiology at high dose, ultra-high dose-rate, single pulse laser-driven proton source}

\author{A. Flacco}
\email{alessandro.flacco@polytechnique.edu}
\author{E. Bayart}
\affiliation{Laboratoire d'Optique Appliqu\'ee, ENSTA Paris, CNRS, Ecole Polytechnique,
Institut Polytechnique de Paris, 91120 Palaiseau, France}
\author{L. Romagnani}
\affiliation{LULI, CNRS, \'Ecole Polytechnique, Institut Polytechnique de Paris, 91120 Palaiseau, France}
\author{M. Cavallone}
\affiliation{Centre de Protonth\'erapie d'Orsay, Department of Radiation Oncology, Campus
Universitaire, Institut Curie, PSL Research University, 91898 Orsay, France}
\author{L. DeMarzi}
\affiliation{Centre de Protonth\'erapie d'Orsay, Department of Radiation Oncology, Campus
Universitaire, Institut Curie, PSL Research University, 91898 Orsay, France}
\affiliation{INSERM LITO 1288, Campus Universitaire, Institut Curie, PSL Research University,
University Paris-Saclay, 91898 Orsay, France}
\author{C. Fouillade}
\affiliation{Institut Curie, INSERM U1021-CNRS UMR 3347, University Paris-Saclay, PSL Research University, Centre Universitaire, Orsay Cedex, France.}
\author{C. Giaccaglia}
\affiliation{Laboratoire d'Optique Appliqu\'ee, ENSTA Paris, CNRS, Ecole Polytechnique,
Institut Polytechnique de Paris, 91120 Palaiseau, France}
\author{S. Heinrich}
\affiliation{Institut Curie, INSERM U1021-CNRS UMR 3347, University Paris-Saclay, PSL Research University, Centre Universitaire, Orsay Cedex, France.}
\author{I. Lamarre-Jouenne}
\affiliation{Laboratoire d'Optique et Biosciences, Ecole Polytechnique, CNRS, Inserm, Institut Polytechnique de Paris, Palaiseau, France}
\author{J. Monzac}
\affiliation{Laboratoire d'Optique Appliqu\'ee, ENSTA Paris, CNRS, Ecole Polytechnique,
Institut Polytechnique de Paris, 91120 Palaiseau, France}
\author{K. Parodi}
\affiliation{Ludwig-Maximilians-Universit\"at M\"unchen, Faculty of Physics, Department of
Medical Physics, 85748 Garching b. M\"unchen, Germany }
\author{A. Patriarca}
\affiliation{Centre de Protonth\'erapie d'Orsay, Department of Radiation Oncology, Campus
Universitaire, Institut Curie, PSL Research University, 91898 Orsay, France}
\author{T. R\"osch}
\author{J. Schreiber}
\author{L. Tischendorf}
\affiliation{Ludwig-Maximilians-Universit\"at M\"unchen, Faculty of Physics, Department of
Medical Physics, 85748 Garching b. M\"unchen, Germany }

\keywords{Laser-driven proton acceleration, ultra-high dose-rate, FLASH, quadrupoles,
dosimetry}

\begin{abstract}
	Laser-driven proton sources have long been developed with an eye on their
	potential for medical application to radiation therapy. These sources are compact,
	versatile, and show peculiar characteristics such as extreme instantaneous dose
	rates, short duration and broad energy spectrum. Typical temporal modality of
	laser-driven irradiation, the so-called \emph{fast-fractionation}, results from
	the composition of multiple, temporally separated, ultra-short dose fractions.\\
	In this paper we present the use  of a high-energy laser system for delivering the
	target dose in a single nanosecond  pulse, for ultra-fast irradiation of
	biological samples. A transport line composed by two permanent-magnet quadrupoles
	and a scattering system is used to improve the dose profile and to control the
	delivered dose-per-pulse. A single-shot dosimetry protocol for the broad-spectrum
	proton source using Monte Carlo simulations was developed. Doses as high as
	\SI{20}{\gray} could be delivered in a single shot, lasting less than \SI{10}{\ns}
	over a \SI{1}{\cm} diameter sample holder, at a dose-rate exceeding
	\SI{e9}{\gray\per\s}.  Exploratory application of extreme laser-driven irradiation
	conditions, falling within the FLASH irradiation
	protocol\cite{vozeninClinicalTranslationFLASH2022}, are presented for in vitro and in vivo irradiation. A reduction of radiation-induced oxidative
	stress in-vitro and radiation-induced developmental damage in vivo were observed,
	whereas anti-tumoral efficacy was confirmed by cell survival assay.
\end{abstract}

\flushbottom
\maketitle
\thispagestyle{empty}

\section*{Introduction}
\label{sec:intro}
Radiation therapy is a cornerstone in cancer management. Apart from X-rays, which
represent the strong majority of treatments, other radiation qualities and disparate
spatial or temporal source parameters are used to match particular therapeutic needs.
Unlike X-rays, protons have a rapid distal dose fall-off and a reduced proximal dose.\\
Laser-driven proton sources have been proposed as a promising alternative to conventional
(cyclotrons, synchro-cyclotrons)
accelerators\cite{zeilDosecontrolledIrradiationCancer2013,masoodCompactSolutionIon2014}.
Lasers, as an energy source, do not require specific radioprotection; moreover ion plasma
sources produce broad spectra, which could offer novel alternative strategies for
producing a spread-out Bragg peak (SOBP), used for the treatment of deep-seated tumors.
As of today, existing laser technologies and explored acceleration strategies do not
provide sufficient kinetic energies for the medical application, although successful
experiment of \emph{in-vitro}\cite{bayartFastDoseFractionation2019,
pommarelSpectralSpatialShaping2017} and surface
\emph{in-vivo}\cite{krollTumourIrradiationMice2022} irradiation made a huge step forward
to demonstrate their practical use in radiation biology.

The recent discovery of the FLASH effect by Favaudon et
al.\cite{favaudonUltrahighDoserateFLASH2014a} renewed the interest around the role of
dose-rate and temporal modality of dose deposition in defining the biological and the
physiological effect of ionizing radiation. Some advantages demonstrated for
\emph{in-vivo} and human irradiation (see Wilson et al.\cite{wilsonUltraHighDoseRate2020}
and references therein for a consistent review) suggest a possible improvement of the
therapeutic window at high dose rate, which motivates the exploration of non-conventional
temporal dose deposition modalities.  Laser-driven particle sources produce short and
bright particle bunches, as each laser pulse independently extracts and accelerates a
short and bright packet.  Owing to the ultra-fast nature of laser pulses at
relativistic intensity, laser-accelerated protons have a duration, at the source,
between few picoseconds and few
nanoseconds\cite{moraThinfoilExpansionVacuum2005,dromeyPicosecondMetrologyLaserdriven2016}.

An ever increasing number of laboratories worldwide are exploring the biological effects
of laser-accelerated protons, where relevant irradiation conditions are met in terms of
particle penetration, irradiated volume and average dose-rate. Such
systems are often limited to an energy-per-pulse of a few to tens of joules and
repetition rates in the Hz range or lower. The total useful  proton charge from such
systems is sufficient to deposit a dose of a few \si{\milli\gray} to a fraction of
\si{\gray} per shot, where a projected target surface of \SI{1}{\cm\squared} is considered. In this
condition, multiple laser pulses are required to produce a target dose within \num{1} to
\SI{10}{\gray}, with a total irradiation time not shorter than a few seconds and up to a
few minutes. This irradiation modality, where the dose is deposited by separate fractions
at ultra-high dose-rate, termed ``fast fractionation'', was shown to produce particular
biological effects in some
conditions\cite{bayartFastDoseFractionation2019,raschkeUltrashortLaseracceleratedProton2016}.

In this paper we present the use of a kJ-class laser to achieve the total target dose
within a single laser-driven proton pulse. A set of permanent-magnet quadrupoles is used
to transport and shape the protons to an irradiation area in air, far from the
interaction point. This technique gives access to doses as high as \SI{20}{\gray}
delivered within a few nanoseconds over a \SI{1}{\cm\squared} \emph{in-vitro} sample.\\
The transport line provides an additional control on the irradiation conditions, such as the
ability to vary the projected charge density in the proton bunch, for performing
dose-escalation experiments. Moreover, the spectral selection operated by the quadrupoles
can be used to produce a uniform depth dose deposition (SOBP) within thicker targets
(up to \SI{600}{\um} of water). \\
In order to correctly apply our particle source to radiobiology experiments, a dosimetry
protocol for the reconstruction of the spectral content of each laser shot is developed
and applied to all of the explored irradiation conditions. \\
A survival assay of human glyoblastoma cell line U87-MG is performed as a confirmation of the
toxicity to cancerous cells of the irradiation protocol. The generation of stress-dependent oxydative stress is studied
in-vitro in healthy (MRC5) and tumoral (U87-MG) cell lines, showing reduced DNA damage
in the healthy cells. This suggests a protective effect of the high-dose-rate irradiation
condition. Evaluation of zebrafish embryos development following irradiation under these
conditions suggest onset of the FLASH effect with short, laser-driven, single
pulse proton irradiation scheme.

\section*{Source and transport}
\begin{figure}[!ht]
	\begin{center}
		\includegraphics[width=9cm]{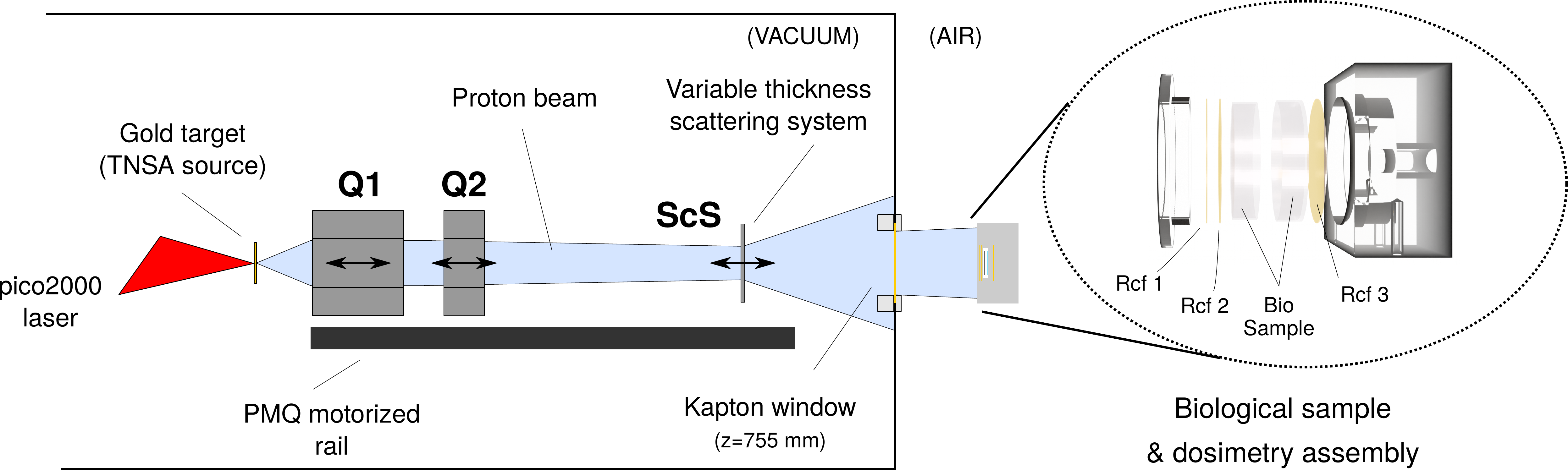}
	\end{center}
	\caption{Scheme of the experimental setup, showing the TNSA proton source, the transport
		line composed by two permanent-magnet quadrupoles (Q1, Q2), and the scattering
	system (ScS), all contained within a vacuum, along with the enclosure for dosimetry and biological sample in air.}
	\label{fig:scheme}
\end{figure}

The experiment was performed in two separate campaigns at the LULI laboratory (\'Ecole Polytechnique, Palaiseau,
France) on the \emph{pico2000} laser system. The picosecond beam (duration
\SI{1.2}{\ps}, energy \SI{100}{\joule} per pulse, $\lambda=\SI{1053}{nm}$) is
focused by a $f/4$ off-axis parabola (PFL \SI{800}{\mm}, d=\SI{180}{\mm}).

Protons are accelerated in the \emph{target-normal sheath acceleration} (TNSA) scheme.
Optimal acceleration conditions, showing both highest charge and cut-off energy, were
determined from previous experimental campaigns\cite{perozzielloLaserdrivenBeamsFuture2016}.  The TNSA
mechanism relies on the energy transfer from the laser-heated electron
fraction to the cold ion plasma, during the plasma expansion in vacuum
(see Macchi 2017\cite{macchiReviewLaserPlasmaIon2017} and references
therein); ion spectrum is thermal with a high-energy cutoff.  In order to
characterize the particle source for magnetic transport design, a set of
stacked radiochromic films (RCF) is used. The stack is installed at $z=\SI{48}{\mm}$
from the interaction point ($z=0$) and irradiated with protons from a single
laser shot. The spectral and spatial features of the accelerated beam are
extracted from the irradiated RCFs following an exponential fitting
routine (described in section 3.1.2 of
Cavallone,~2020\cite{cavallone:tel-03085030}) which solves with higher
precision the cutoff energy and the overall spectral shape.  The analysis
procedure also enables a rough estimate of the total available charge at
the source.

\begin{figure}[!ht]
	\begin{center}
		\setlength{\unitlength}{1cm}
		\begin{picture}(8, 10)(0,0)
			\put(0,5){\includegraphics[width=8.5cm]{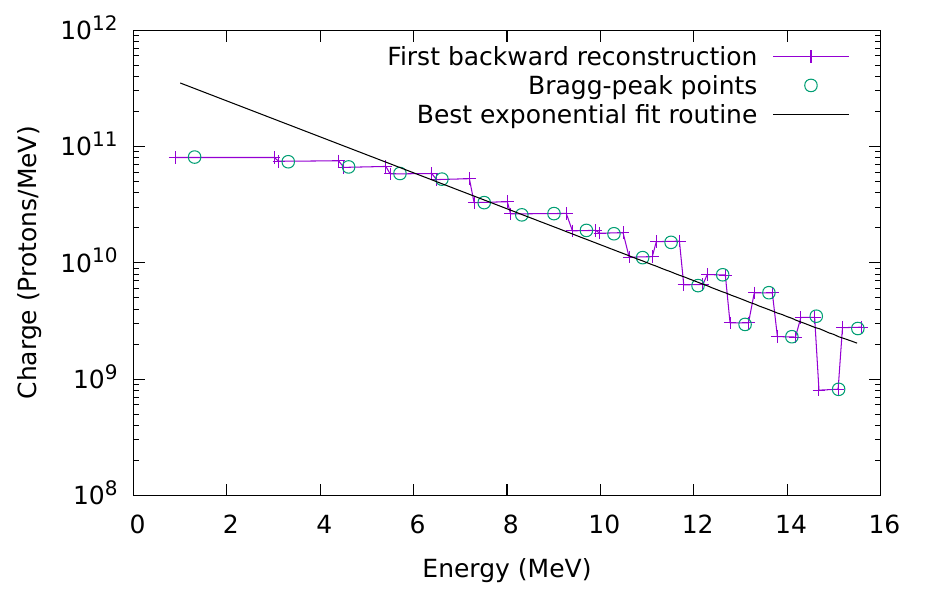}}
			\put(0.3,0){\includegraphics[width=8cm]{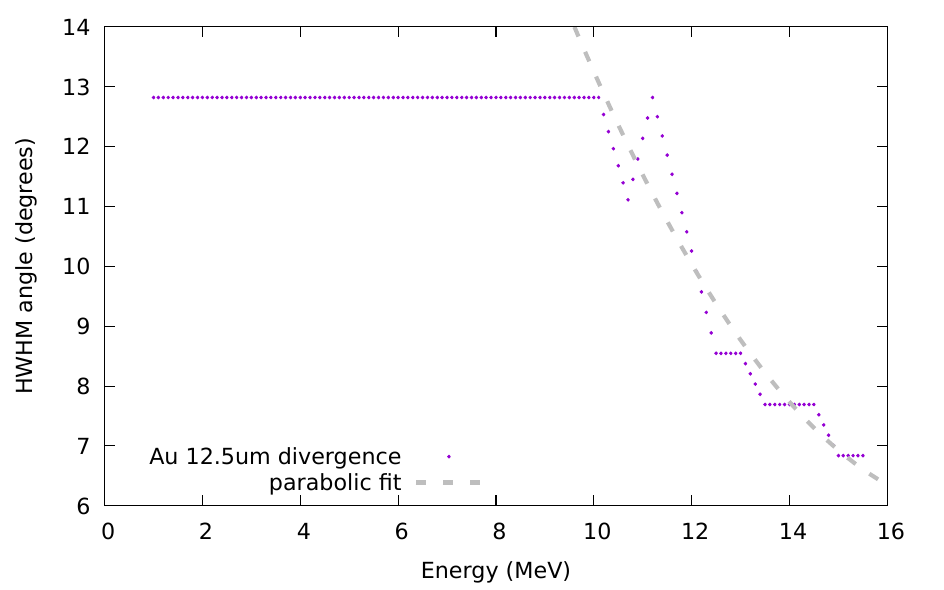}}
			\put(0,9.5){\textbf{(a)}}
			\put(0,4.8){\textbf{(b)}}
		\end{picture}
		\caption{Spectral charge \textbf{(a)} and spectral divergence
			\textbf{(b)} on a single calibration shot,
			reconstructed from a stack of 25 HD810 type foils. Foils \#\num{1}
			through \#\num{4} are
			saturated; foils \#\num{20} through \#\num{25} were not exploitable for the fitting procedure and
only used to assess the cut-off energy.}
	\label{fig:source-parameters}
	\end{center}
\end{figure}

A calibration shot to assess the source charge and spectral features is illustrated in
\figref{fig:source-parameters}a for a \SI{12.5}{\um} thick gold target. The spectrum
exhibits the expected exponential profile, with a cutoff at \SI{19.6}{\MeV}. Radiochromic
films \#\num{1}
through \#\num{4} were saturated. Different spectral components exhibit a decreasing
divergence at increasing energy (\figref{fig:source-parameters}b). It's interesting to
note that protons with energy ${\rm E}<\SI{10}{\MeV}$ are emitted with constant
divergence, producing a rather sharp boundary on the radiochromic films deeper in the stack. At
higher energies, ${\rm E}>\SI{10}{\MeV}$ the beam divergence decreases following a parabolic profile. According
to the general calibration function of the HD810 film, the total charge within the
exponential fit for the presented shot exceeds \SI{150}{\nano\coulomb}.

\subsection*{Transport configuration}
The transport line is designed to produce a controlled irradiation to a plane outside the
experimental chamber. Air-vacuum separation is guaranteed by a \SI{75}{\um} thick Kapton
film, situated at $z=\SI{772}{\mm}$ from the interaction point, whereas the irradiation
plane lays at $z=\SI{827}{\mm}$ (\figref{fig:scheme}). The transport line is composed
by two permanent-magnet quadrupoles and a variable-thickness scattering filter.
Quadrupoles \textbf{Q1} and \textbf{Q2} are respectively $d_1=\SI{40}{\mm}$ and
$d_2=\SI{20}{\mm}$ long, and have field gradients of $g_1=\SI{332}{\tesla\per\meter}$ and
$g_2=\SI{322}{\tesla\per\meter}$. Quadrupoles construction and calibration is detailed in
Rösch et al.\cite{roschCharacterizationPermanentMagnet,roschOptimizationPermanentMagnet2021}. Both
quadrupoles have an inner bore of \SI{1}{\cm} diameter; considering the shortest possible
distance between quadrupole Q1 and the proton source,
$\Delta z=\SI{48}{\mm}$, the bore sets a maximum angular acceptance of $\theta_{max}
= \SI{100}{\milli\radian}$. 
The scattering system (\textbf{ScS}) is composed by four different selectable aluminum
filters of varying thickness to be
inserted between the quadrupole Q2 and the Kapton window, to provide diffusion of the proton beam,
improving divergence and transverse uniformity. 
All elements on the transport line are motorized and equipped with optical encoders,
enabling reliable positioning even in presence of strong magnetic forces between the
elements.\\

The role of the transport line is to control the particle density and, consequently, 
the deposited dose at the target plane, far from the laser-plasma interaction point. During the design step
the deposited dose at the irradiation plane is simulated by Monte Carlo methods (Geant4
toolkit) for different beam line configurations.
The source is modeled as being purely thermal with a high energy cutoff 
and no angular dependence. This assumption is based on the small effective aperture of the
first quadrupole, which acts as a filter on the broad divergent source. The largest
accepted half-width angle ($\theta_{max} = \SI{6}{\degree}$) is, according to measurements in
\figref{fig:source-parameters}, smaller than the source divergence at the cutoff energy.
For this reason we decided to drop the angular dependence in the source modelling; the
useful spectral charge is then represented analytically as:
\begin{equation}
	Q\left( E \right) \id{E}= \frac{Q_0^{*}}{E_0} \exp \left[ - E /E_0 \right] \id{E},
	\label{eq:spectrum}
\end{equation}
where E is the proton kinetic energy, E$_0$ the spectral distribution temperature and Q a charge. The term Q$_{0}^{*}$ (measured
in nC$^*$ from now on to avoid confusion) indicates that the integrated charge of the
source term expressed in equation \eqref{eq:spectrum} differs from the total charge at the
laser-plasma source, as it is limited to the charge emitted within the first quadrupole
input acceptance angle (see Monte Carlo simulation section in Methods for
further details).\\

\begin{figure}[!ht]
	\begin{center}
		\setlength{\unitlength}{1cm}
		\begin{picture}(8,14.5)(0,0)
			\put(0,9.5){\includegraphics[width=7cm,page=2]{MC-transport-dosemaps.pdf}}
			\put(0,14){\textbf{(a)}}

			\put(0,8.2){\textbf{(b)}}
			\put(0.3,3.5){\includegraphics[width=6.5cm,page=1]{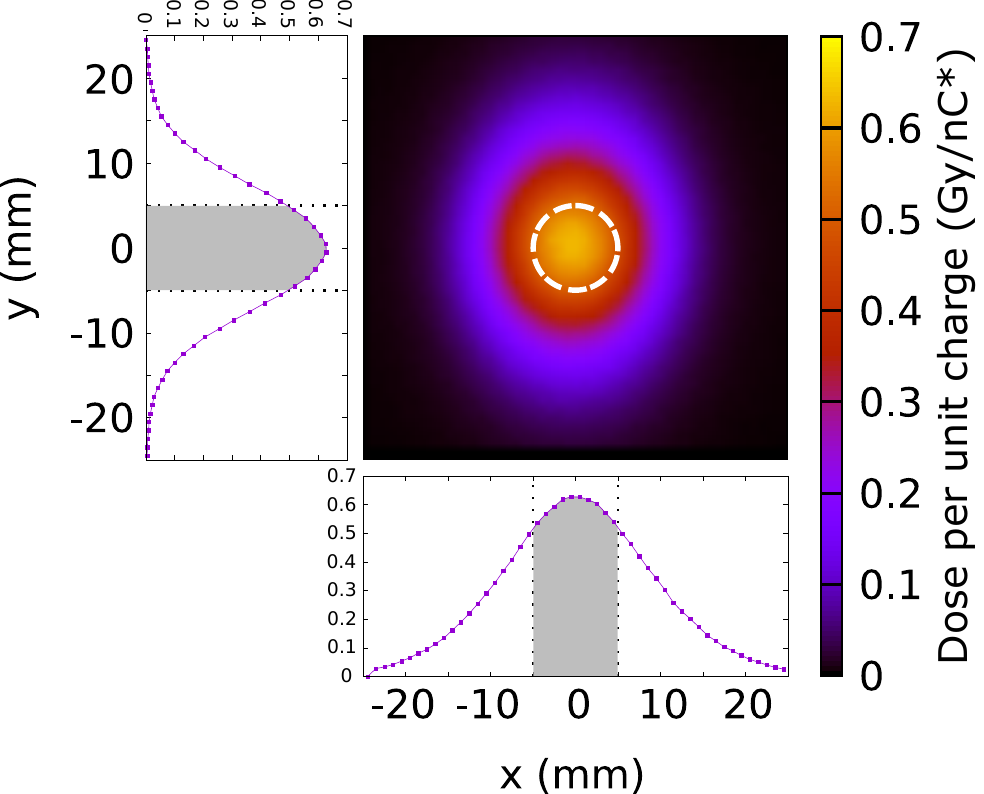}}
			\put(0,3.2){\textbf{(c)}}
			\put(0,0){\includegraphics[width=8cm,page=1]{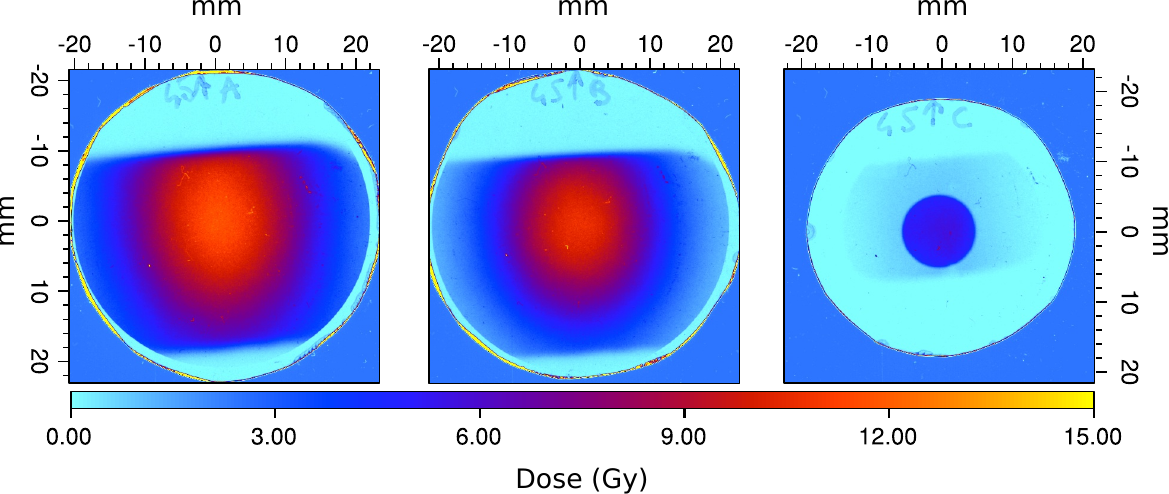}}
		\end{picture}
	\end{center}
	\caption{Deposited dose in the water sample and transverse uniformity at the
		irradiation plane for thin biological samples ($\SI{20}{\um}$ water) in
		the dose escalation (DE) configuration. \textbf{(a)} Simulated dose per
		unit collected charge for varying distance between Q2 and ScS, with a
		\SI{70}{\um} scattering filter. \textbf{(b)} Simulated dose map
	corresponding to a dose escalation configuration with filter at $\Delta z\left[
ScS-Q2 \right]=\SI{200}{\mm}$. \textbf{(c)} Experimental dose map recorded with the sample
holder in place in the same transport configuration as \textbf{(b)}.}
	\label{fig:dose-escalation}
\end{figure}

Two sets of configurations are designed for the particle transport. A first condition (termed \emph{dose
escalation}, DE) is studied for \emph{in-vitro} irradiation. The surface to be irradiated
is set to have a diameter of \SI{1}{\cm}, centered on axis on a Lumox capsule and the
biological sample represented by a \SI{20}{\um} thick water volume. In this condition
the quadrupole Q2 is set at \SI{20}{\mm} from the first quadrupole. 
In this configuration the scattering filter holds a \SI{70}{\um} aluminum foil and is moved to vary the particle density
at the irradiation plane, hence the deposited dose within the projected surface (\figref{fig:dose-escalation}).

\begin{figure}[!ht]
	\begin{center}
		\setlength{\unitlength}{1cm}
		\begin{picture}(8.5, 13.5)(0,0)
			\put(0,13){\textbf{(a)}}
			\put(1,9.2){\includegraphics[height=45mm,page=4]{MC-transport-dosemaps.pdf}}

			\put(0,8){\textbf{(b)}}
			\put(0.2,3){\includegraphics[width=7cm,page=3]{MC-transport-dosemaps.pdf}}

			\put(0,2.8){\textbf{(c)}}
			\put(0,0){\includegraphics[width=8.5cm]{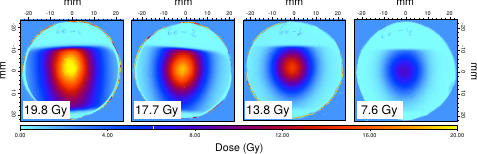}}
			
		\end{picture}
	\end{center}
	\caption{Deposited dose in a thick sample (\SI{600}{\um} water) in the ZF configuration. \textbf{(a)} Dose per unit input charge at
		varying depth in water with $\Delta z\left[
		ScS-Q2 \right]=\SI{250}{\mm}$ and a scattering filter of
		\SI{50}{\um} aluminum. \textbf{(b)} Simulated 2D map of integrated dose through the
		entire sample. \textbf{(c)} Experimental dose maps recorded on a stack of
		EBT-XD type radiochromic films (water equivalent film thickness:
	\SI{385}{\um}) with the transport line. Inset doses
refer to the ROI marked in panel \textbf{(b)}.} \label{fig:dose-depth}
\end{figure}

An additional configuration was studied for \emph{in vivo} irradiation (Zebrafish
	embryos, ZF). In this condition the target dose at the sample is set to be as close as possible to
a well defined condition (\SI{8}{\gray}, as in Bourhis et
al.\cite{bourhisClinicalTranslationFLASH2019}) while keeping the maximum in-depth
(SOBP) uniformity. This irradiation condition was obtained with a scattering filter
of \SI{50}{\um} positioned at a distance of \SI{250}{\mm} past the quadrupole Q2; in
this configuration a smaller profile and a higher dose is produced at
the irradiation plane, as depicted in \figref{fig:dose-depth}. The corresponding
region-of-interest (ROI) is here limited to a diameter of \SI{5}{\mm}.
.

\section*{Single pulse irradiation and dosimetry}
In our experimental condition, the target dose is deposited by the charge driven by a
single laser pulse, within a time shorter than \SI{10}{\ns}. During irradiation it is not
possible to monitor or regulate the deposited dose with a monitor chamber, as performed
during previous
experiments\cite{pommarelShapingControlLaserproduced2017}. 
Spatially resolved dose maps
are recorded by calibrated EBT3-XD radiochromic films as a reference to
the dose in the target sample. However, the width of the proton spectrum and the low
overall kinetic energy in the beam make the energy loss in the RCF itself non negligible.
Consequently, not only the dose at the RCF differs from the dose at the target
sample, but the sole presence of the RCF does modify the target irradiation conditions.
The shot-to-shot variation of the laser-driven proton source parameters (energy, charge)
we observed did not enable to establish one irradiation condition within an acceptable
error margin beforehand. Laser-driven proton acceleration in the TNSA scheme is in fact
very sensitive to the total energy and to the laser-temporal contrast, which are beyond
the user control.\\
In our case, irradiation driven by a single laser pulse, hinders the ability to define an
average spectrum for calibrating the response radiochromic films on the irradiation line.

\subsection*{TNSA Proton source reconstruction}

\begin{figure}[!ht]
	\begin{center}
		\setlength{\unitlength}{1cm}
		\begin{picture}(6, 13)(0,0)
			\put(1,9.5){\includegraphics[width=4.5cm,page=2]{dosi-scheme-mult.pdf}}
			\put(0.5,4.5){\includegraphics[width=6cm]{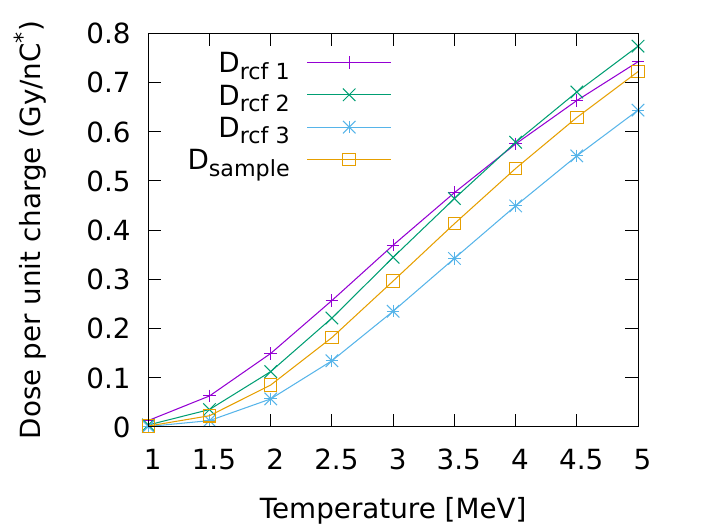}}
			\put(0.5,0){\includegraphics[width=6cm]{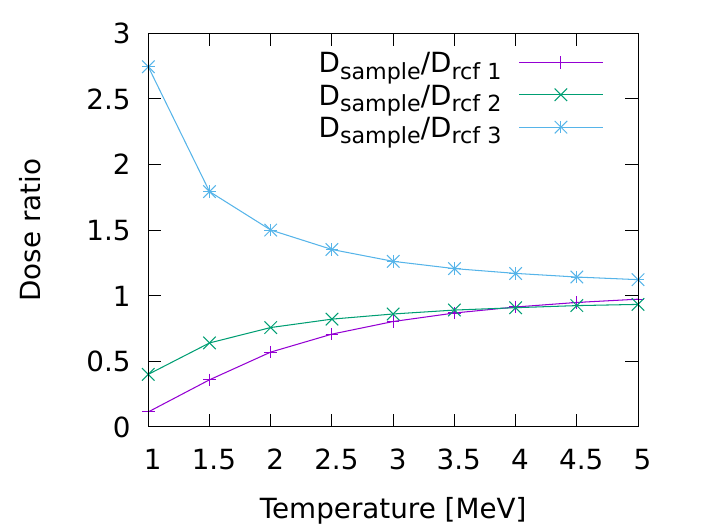}}
			\put(0,12.5){\textbf{(a)}}
			\put(0,8.5){\textbf{(b)}}
			\put(0,4){\textbf{(c)}}
		\end{picture}
	\end{center}
	\caption{Effect of the spectral temperature on dosimeters and biological sample in
		the \emph{in-vitro} irradiation setup on the condition depicted in
		\figref{fig:dose-escalation} ($\Delta z\left[ ScS-Q2
		\right]=\SI{200}{\mm}$). (a) Exploded view of the in-vitro holder, showing
		reference planes in the
		irradiation setup. (b) Simulated dose per unit charge as a function
		of the spectral temperature. (c) Simulated dose ratio between
		biological sample and each of the superposed RCF in the irradiation setup.}
\label{fig:dose-temperature}
\end{figure}

Spectrum and charge changes at the source result in variations of the total deposited
dose, as well as in the dose distribution among different planes in the irradiated volume
(see for example the in-vitro holder structure \figref{fig:dose-temperature}a).
One fundamental assumption on the shot-to-shot behaviour of the proton source is the conservation of
its exponential nature, which is well justified for the TNSA acceleration mechanism.
Under this hypothesis, shot-to-shot parameter change can be considered to affect the only two free
parameters in the spectral distribution \eqref{eq:spectrum}, namely the total charge Q$_{0}^{*}$ and the spectral temperature E$_0$. 
A third parameter known to depend on laser conditions is the  cutoff energy, E$_{high}$,
which is usually defined as a reference parameter for laser-driven proton acceleration
conditions.
For our purposes, a simple rejection criterion can be set for those shots whose cutoff
energy is not sufficient to penetrate the irradiated target stack. Minor
variations can be neglected because, according to Monte Carlo modelling of the irradiation
line, the highest portion of the spectrum contributes to a
lesser extent to the total deposited dose, with respect to the central portion of the
spectrum. For these reasons we consider the cutoff energy constant throughout our analysis.

At least two (although often three of them) RCF were used in most
experimental events, inserted before and
after the biological sample during irradiation (see panel \textbf{(a)} in \figref{fig:dose-temperature}).
Deposited dose in all of the RCFs and the irradiated target is simulated for the
corresponding
beam-line configuration and sources with a varying temperature E$_0$ of the thermal
spectrum between \SIrange{1}{5}{\MeV}.
Following previous considerations and according to equation~(\ref{eq:spectrum}), total charge
Q$_0^*$ and spectral temperature E$_0$ can be extrapolated from the Monte Carlo
simulation which would correctly predict the dose ratio between the superposed RCF.

\begin{figure}[!ht]
	\begin{center}
		\includegraphics[width=10cm]{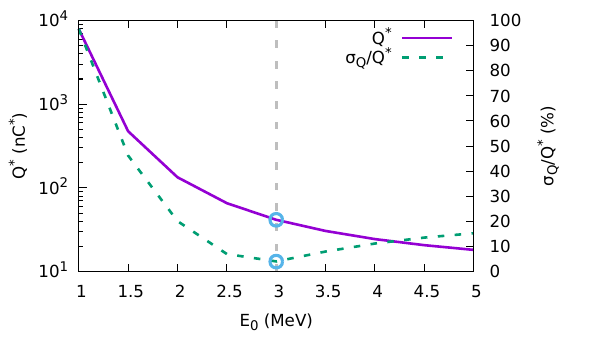}
	\end{center}
	\caption{Example of charge extrapolation for one experimental event (one laser
		shot). Average charge over one set of radiochromic films is plotted at different
	source temperatures; the associated error, is used to determine the best fitting
	temperature. $\sigma_{E_{0}}$ is minimized at $\mathrm{E}_{0}=\SI{3}{\MeV}$ with an error of
\SI{4}{\percent}, which points to a charge of \SI{41.6}{\nC^{*}}.}
	\label{fig:dosi-stat}
\end{figure}

Simulated deposited dose is
scaled to the measured dose on radiochromic films for all the temperatures in the set.
The corresponding charge, averaged for the available RCFs in a single shot, can be then
expressed as:
\begin{eqnarray}
	\overline{Q_{E_{0}}^{*}} &=& \frac{1}{N} \sum_{i}^{N} Q_{i,E_{0}}^{*} = \frac{1}{N}\sum_{i}^{N}
	\frac{D_{i}^{(rcf)}}{D_{i,E_{0}}^{(sim)}}\\[5mm]
	\sigma_{E_{0}} &=& \frac{1}{N} \sqrt{ \sum_{i}^{N} \left(Q_{i,E_{0}}^{*} -
			\overline{Q_{E_{0}}^{*}}
	\right)^{2}, }
	\label{eq:Q-def}
\end{eqnarray}
where $D_{i}^{(rcf)}$ is the experimental integrated dose on the i-th film in the target
and $D_{i,k}^{(sim)}$ is the simulated dose at the spectral temperature E$_0$ on
the i-th film. \figref{fig:dose-temperature}a,b show, as an example, the relationship
between doses at different planes for the case of an experimental shot in the dose
escalation configuration with filter at $\Delta z_{\left[ ScS-Q2 \right]}=\SI{200}{\mm}$.
From the analysis of the calculated dose ratios in the different simulations, it is
possible to set the matching temperature by minimizing the standard deviation of the
extrapolated charge. As an example \figref{fig:dosi-stat} depicts the
minimization of the ratio $\sigma_{Q}/Q^*$ in the case of a test shot, which points to a
temperature of $\mathrm{E}_0 = \SI{3}{\MeV}$. The corresponding value of
$\overline{\mathrm{Q}_{\SI{3}{\MeV}}^*} =
\SI{41.63}{\nano\coulomb^*}$ is then used to rescale the simulation and assess the dose in
the biological sample. The obtained values are shown in the table \ref{table:dose-test}.

\begin{figure}[!ht]
	\begin{center}
		\begin{tabular}[c]{|l|c|c|c|c|}
			\hline
			Position in the stack & {RCF 1} & {RCF 2} & \textbf{Sample} & {RCF 3} \\
			\hline
			Dose (\si{\gray}) & \num{16.2} & \num{13.7} & \textbf{\num{12.34}} & \num{9.7} \\
			\hline
		\end{tabular}
	\end{center}
	\caption{Example of dose reconstruction of a reference shot, showing
		measured dose within the ROI on the three radiochromic films and the
	dose at the sample extrapolated through Monte Carlo simulation.}
	\label{table:dose-test}
\end{figure}

\begin{figure}[!ht]
	\begin{center}
		\setlength{\unitlength}{1cm}
		\begin{picture}(8, 11)(0,0)
			\put(0,5.5){\includegraphics[width=7.8cm]{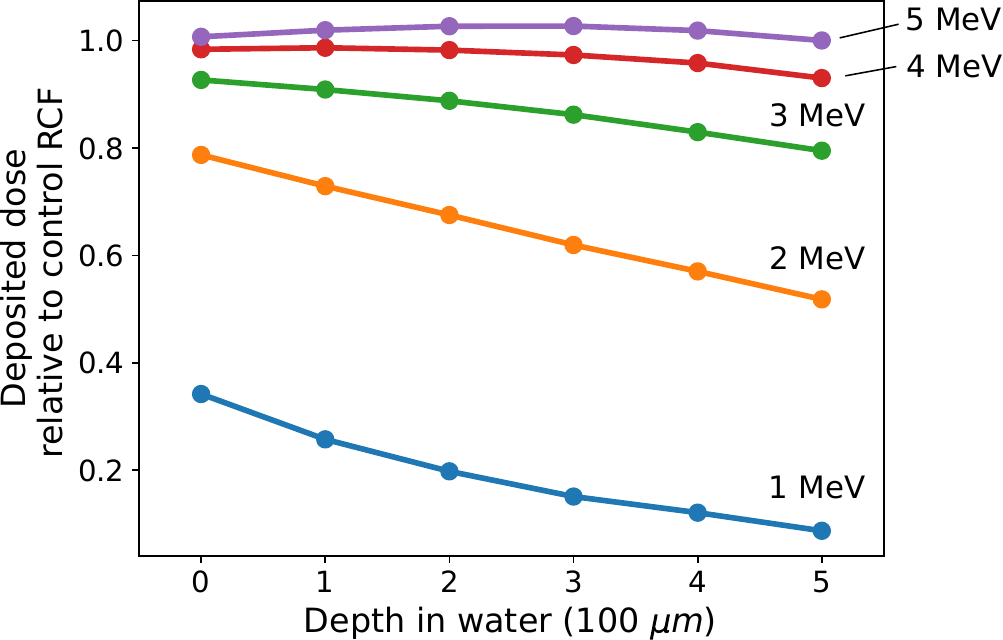}}
			\put(0.25,0){\includegraphics[width=7.2cm]{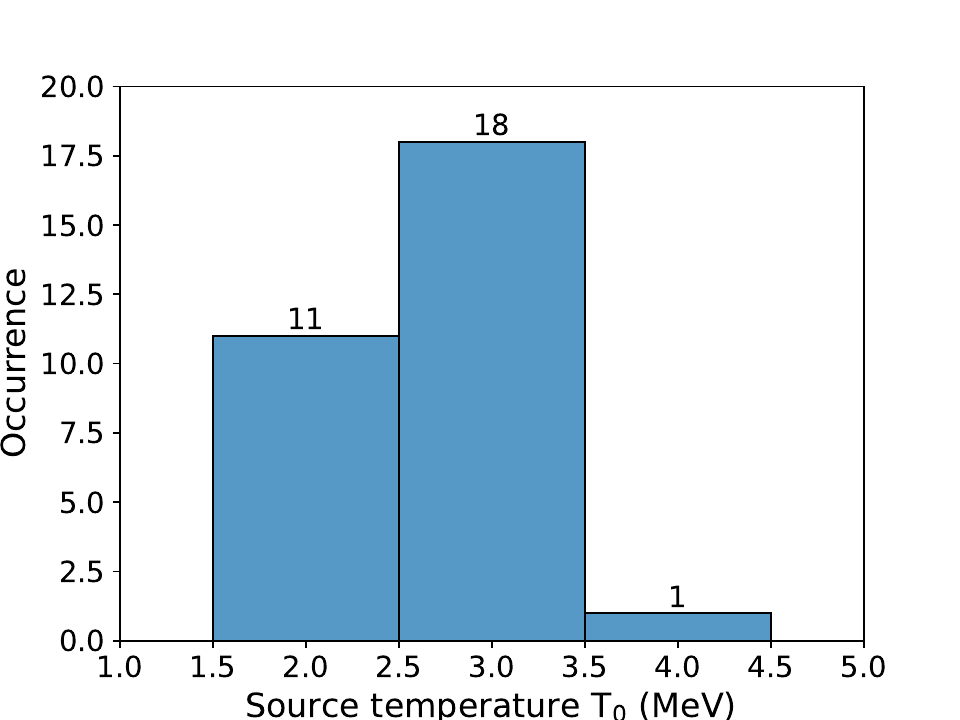}}
			\put(8,11){x}
			\put(0,6){\bf (a)}
			\put(0,0.5){\bf (b)}
		\end{picture}
		\caption{(a) Simulated depth-dose distribution normalized to a single RCF
		foil dose on top of \SI{600}{\um} water target for E$_0$ ranging between
	\SI{1}{\MeV} and \SI{5}{\MeV}. (b) Observed matching temperature during all of the
analysed laser shot where more than one RCF was present on top of the irradiated sample.}
		\label{fig:fish-uniformity}
	\end{center}
\end{figure}

In some of the experimental conditions only one RCF foil was used in front of the
biological target. This was the case for fish irradiation, where we aimed at preserving the best
possible SOBP uniformity; adding a second radiochromic film would have reduced the penetration
depth of the proton beam. Furthermore the actual design of the ZF holder does not allow the insertion
of a back RCF film (see \figref{fig:zf-mount} for details).\\
In these cases only one experimental point is available for assessing the ratio between
the deposited dose and the RCF reading, hence it is not possible to determine the
parameters E$_0$ and Q$_0^{*}$ in a unique way using the previously described protocol. In
order to assess the ratio between the RCF reading and the deposited dose we decided to use the temperature
distribution from the in-vitro events,
shown in \figref{fig:fish-uniformity}b as a statistical weight to the depth-dose curve
and ratio to RCF depicted in \figref{fig:fish-uniformity}a.\\
Following this method, the quantities \textit{sample-dose-to-RCF} and \textit{depth-dose-error} are calculated as
\begin{align}
	\frac{D_{sample}^{exp}}{D_{rcf}^{exp}} &= \sum_{E_0} \left( \frac{n_{E_0}}{n_{tot}}\right) \left( \frac{D_{avg}}{D_{rcf}}
	\right)_{E_0} = 0.78\\[5mm]
	\frac{\sigma_{sample}^{exp}}{D_{sample}^{exp}} &= \sum_{E_0} \left( \frac{n_{E_0}}{n_{tot}}\right) \left(
		\sigma_{depth}
	\right)_{E_0}  = 0.09
	\label{eq:fish-dose-renorm}
\end{align}
where the weights $n_{E_{0}}/n_{tot}$ (spectrum temperature occurrence probability) are
represented in \figref{fig:fish-uniformity}.

\begin{widetext}
\begin{figure*}[!ht]
	\begin{center}
		\begin{tabular}[c]{|c|c|c|c|c|}
			\hline
			Temperature E$_0$ (MeV) & Dose $D_{avg}$ (Gy/nC$^*$) & Depth-dose error
			(relative) & RCF Dose $D_{rcf}$ (Gy/nC$^*$) & Dose ratio
			$\frac{D_{avg}}{D_{rcf}}$ \\
			\hline
			\num{1.0} & \num{3.0e-4} & \num{1.5e-4} (\SI{48.9}{\percent}) & \num{1.58e-2} & \SI{19.25}{\percent} \\
			\hline
			\num{2.0} & \num{0.114} & \num{17.6e-3} (\SI{15.4}{\percent}) & \num{0.176} & \SI{64.98}{\percent} \\
			\hline
			\num{3.0} & \num{0.382} & \num{21e-3} (\SI{5.7}{\percent}) & \num{0.44} & \SI{86.8}{\percent} \\
			\hline
			\num{4.0} & \num{0.66} & \num{14.8e-3} (\SI{2.2}{\percent}) & \num{0.68} & \SI{96.9}{\percent} \\
			\hline
			\num{5.0} & \num{0.908} & \num{9.6e-3} (\SI{1.0}{\percent}) & \num{0.89} & \SI{101.6}{\percent} \\
			\hline
		\end{tabular}
		\caption{Parameters for depth-dose distribution at varying temperature, as in
		\figref{fig:fish-uniformity}-A}
		\label{table:fish-uniformity}
	\end{center}
\end{figure*}
\end{widetext}

\section*{Application to radiation biology}
The irradiation beam-line was used for irradiation on several in-vitro and in-vivo
biological targets, in
order to verify its functional hypothesis, provide a rough validation of the dosimetry
protocol and start to explore this novel irradiation condition.\\
In the following, radiation qualities are indicated as \textbf{SP-LDP} for ``single-pulse
laser-driven protons'', \textbf{FF-LDP} for ``fast-fractionated laser-driven protons''
(indicating a dose deposited by multiple separated fraction at ultra-high instantaneous
dose-rate) and \textbf{CAP} for ``conventional-accelerated protons''.

\subsubsection*{Cell survival assay after exposure to laser-driven protons (LDP) at pico2000}

\begin{figure}[!ht]
	\begin{center}
		\includegraphics[width=0.50\textwidth]{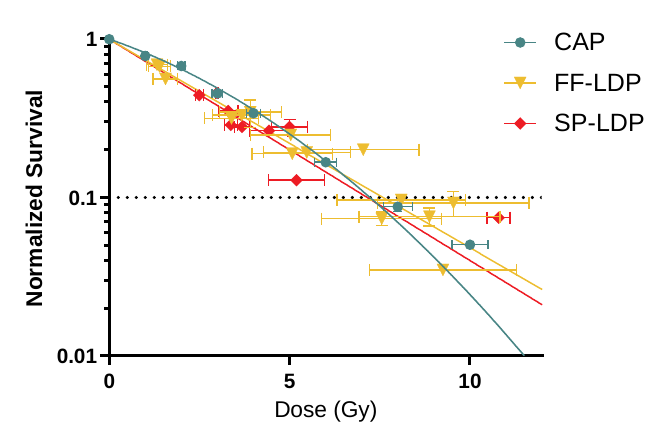}
	\end{center}
	\caption{Cell survival dose-response of U87-MG cell line. Normalized cell
		survival resulting from exposure to increasing doses of CAP, FF-LDP and SP-LDP in
		U87-MG cells. Each data point represents the mean and standard deviation (SD) of three
		replicates obtained at least with three independent experiments. Survival curves were
		generated following the linear quadratic model (R squared values were \num{0.9756},
	\num{0.9773} and \num{0.9824} for CAP, FF-LDP and SP-LDP respectively).}
	\label{fig:u87-surv}
\end{figure}

The impact of SP-LDP is evaluated on the highly resistant
glioblastoma cells of the line U87-MG, through a cell-survival assay. Cells are prepared
as in a previous experimental campaign\cite{bayartFastDoseFractionation2019} where the
effect of laser-pulse pacing in a fast-fractionation irradiation modality was explored. It's
worth noting that U87-MG showed no sensitivity to the average dose-rate at fixed dose
within the range explored during previous experiments (dose-per-fraction:
\SI{0.7}{\gray}, delay between fractions: $\SI{2}{\s} \le \Delta t \le \SI{60}{\s}$). \\
U87-MG cells were irradiated in single pulse modality with doses ranging from \num{2.5} to
\SI{10.8}{\gray} and the resulting dose-response survival curves obtained from
non-clonogenic survival assays (see Methods). \figref{fig:u87-surv} show a superposition
between novel SP-LDP points and those already published in Bayart et al., 2019
\cite{bayartFastDoseFractionation2019}. The D$_{10}$ values resulting from a
linear-quadratic model fit do show good
agreement between the three experiments, confirming the toxicity of our irradiation
conditions on cancerous cells and the validity of the dose escalation scheme. The set of 
$D_{10}$ values across the two experiments is summarized in
Table~\ref{table:dose-comparison}, along with average and instantaneous dose-rate
(respectively $\overline{\dot{\mathrm{D}}}$ and ${\dot{\mathrm{D}}}$) and the $R^2$ value
of the fit.

\begin{figure}[!ht]
	\begin{center}
		\renewcommand{\arraystretch}{1.6}
		\begin{tabular}{|l|c|c|c|}
			\hline
			Radiation & CAP & FF-LDP & SP-LDP \\ 
			\hline
			D$_{10}$ (Gy) & \color{gray} \num{7.11 \pm 0.16} & \color{gray} \num{7.47 \pm 0.32} & \num{7.13 \pm 0.29} \\
			\hline
			$\overline{\dot{\mathrm{D}}\,\,(\si{\gray\per\second})}$ &\color{gray}
			\num{5e-2} &\color{gray} $0.3$ - $0.7$
			&\num{e8} - \num{e9}  \\
			${\dot{\mathrm{D}}\,\,(\si{\gray\per\second})}$ &\color{gray}
			\num{5e-2} &\color{gray}  \num{1.5e8} &
			\num{e8} - \num{e9}\\
			\hline
			$R^2$ & \color{gray} \num{0.9756} & \color{gray} \num{0.9773} & \num{0.9824} \\
			\hline
		\end{tabular}
		\caption{Comparison of doses giving \SI{10}{\percent} of cell survival
			(D$_{10}$), average ($\overline{\dot{\mathrm{D}}}$) and instantaneous
			($\dot{\mathrm{D}}$) dose-rates from CAP, FF-LDP and SP-LDP. Mean D$_{10}
			\pm$ SEM extracted from curves obtained in \figref{fig:u87-surv} are
		reported. Grayed values from Bayart et al.\cite{bayartFastDoseFractionation2019}} 
		\label{table:dose-comparison}
	\end{center}
\end{figure}

\subsubsection*{Oxidative stress-dependent DNA damage in healthy and tumoral
cell lines after exposures to SP-LDP.}

In the described conditions, the total dose is deposited in a time shorter than
\SI{10}{\ns}, at a dose-rate exceeding \SI{e8}{\gray\per\second}.  These conditions are
several orders of magnitude shorter and more intense than those indicated as a threshold
for FLASH effect in Bourhis et al, 2019\cite{bourhisClinicalTranslationFLASH2019}. In this single-pulse
condition, dose deposition happens in a temporal span faster than the homogeneous chemical
step\cite{montay-gruelLongtermNeurocognitiveBenefits2019}.  Although the associated
mechanistics is not yet known, FLASH effect has been associated to the reduction of
oxidative stress in healthy tissue. 

\begin{figure}[!ht]
	\begin{center}
		\includegraphics[width=6cm]{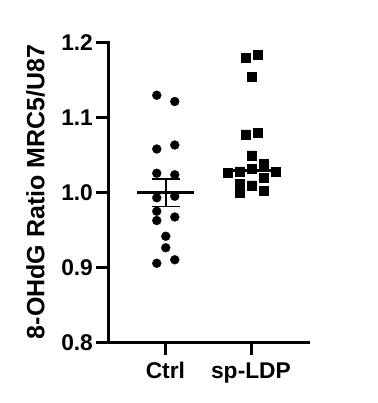}
	\end{center}
	\caption{Oxidative stress-dependent DNA damage in SP-LDP irradiation.
		8-Hydroxy-2'-deoxyguanosine (8-OHdG) production ratio between MRC5 and
		U87-MG cell lines, for both non-irradiated and SP-irradiatied cells,
		\SI{1}{\hour} post irradiation. Mean doses deposited were
		\SI{3.1 \pm 0.50}{\gray} and \SI{2.37 \pm 0.59}{\gray} in MRC5 and U87-MG cells
		respectively.  Mean ratio and SEM are represented, $p = 0.0076$,
		Kolmogorov-Smirnov test.
	}
	\label{fig:oxy-mrc5-u87}
\end{figure}

To address the question whether SP-LDP could have a different effectiveness to generate
oxidative stress-dependent DNA damage, we aimed to measure 8-Hydroxy-2'-deoxyguanosine
(8-OHdG). 8-OHdG is one of the most widely studied oxidized metabolites and is considered
a biomarker for oxidative damage of
DNA\cite{kasaiAnalysisFormOxidative1997,beckmanOxidativeDecayDNA1997}. Human healthy
fibroblasts MRC5 and U87-MG cells were exposed to a target dose of \SI{2.5}{\gray} of
SP-LDP (\SI{3.1 +- 0.50}{\gray} and \SI{2.37 +- 0.59}{\gray} in MRC5 and U87-MG cells
respectively according to post-irradiation dosimetry). The amount of 8-OHdG was determined
in cells harvested one hour later  and analyzed by Elisa assay (see Methods). Using this
detection method, the amount of 8-OHdG detected is inversely related to absorbance.
\figref{fig:oxy-mrc5-u87} shows the ratio of the values obtained with MCR5 cell line
out of those of U87-MG. At the basal level, MRC5 and U87-MG cells present similar amounts
of 8-OHdG, indicated by a ratio close to one. However, following irradiation with SP-LDP,
8-OHdG ratio increase significantly ($p = 0.0076$, Kolmogorov-Smirnov test). These result 
suggests that, following exposure to ultra-high dose-rate protons, the glioblastoma cell
line model receives a higher fraction of DNA damage due to oxidative stress than he
healthy cell model. 
This is the first time that a link between DNA damage, oxidative
stress and the differential response between healthy and tumor cells is observed. It may be
also be an indication of an \emph{in vitro} sparing effect at ultra-high dose-rate,
similar to FLASH irradiation protocols.

\subsubsection*{Zebrafish embryo development evaluation as preclinical \emph{in-vivo} model for
radiation research at pico2000}
\begin{widetext}
	\begin{figure*}[!ht]
		\begin{center}
			\setlength{\unitlength}{1cm}
			\begin{picture}(17.5, 4.7)(0,0)
				\put(0,1){\includegraphics[width=50mm]{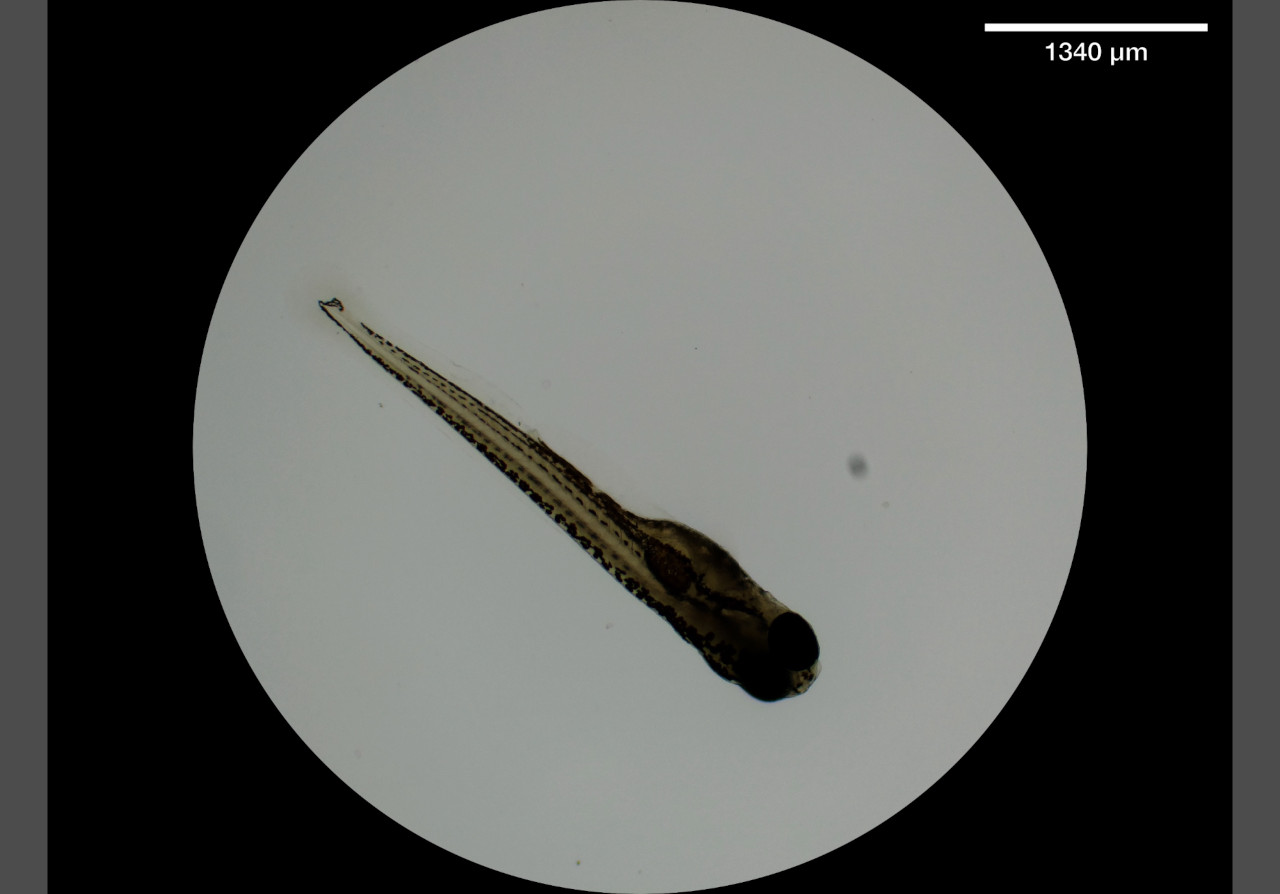}}
				\put(5.5,0){\includegraphics[width=65mm]{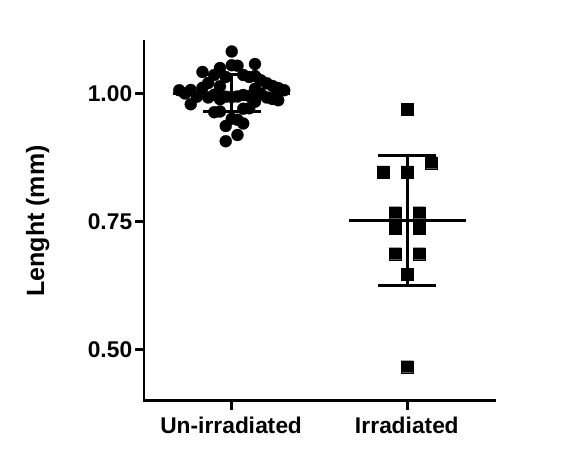}}
				\put(12.5,1){\includegraphics[width=50mm]{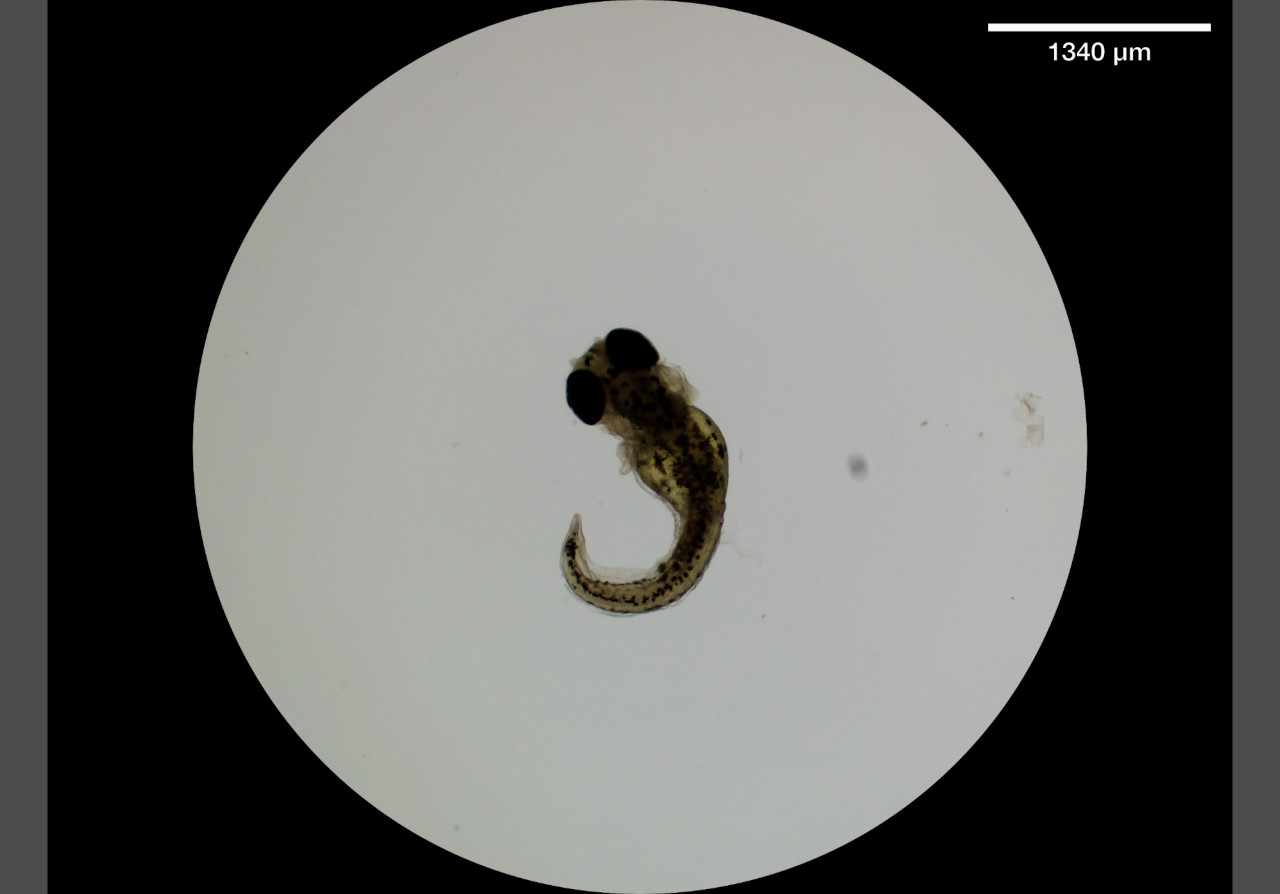}}
			\end{picture}
		\end{center}
		\caption{Zebrafish embryo development evaluation. Zebrafish embryos were
			irradiated \SI{4}{hours} post-fertilization, length of animals was measured 5 days
			post-irradiation. Example of non-irradiated (left) and of irradiated
			(right) fish, exhibiting spine curvature and malformations. (Center) Length of animals
			were expressed as ratios of the measured values for each condition out of the mean length
			value of non-irradiated animals. The plotted data correspond to embryos from three
			independent irradiations, from three independent fertilizations and days of the
			experimental campaign. Each data point represents one living animal. Mean dose deposited
			was \SI{9.71 +- 1.37}{\gray}.  Mean ratio and SEM are represented, $n=3$, 7 embryos per
		group, $p < 0.0001$ (unpaired t-test).}
		\label{fig:zf-length}
	\end{figure*}
\end{widetext}

Radiation-induced toxicity reduction at constant dose and higher dose-rate (FLASH effect) have been exclusively observed with \emph{in vivo} models (for review see Vozenin et
al, 2022\cite{vozeninClinicalTranslationFLASH2022}) including mice, cats, dogs,
pigs and zebrafish (\emph{Danio rerio}). Aiming at a validation of \emph{in vivo}
irradiation of laser-driven protons, zebrafish do represent a good
candidate. 
Zebrafish (ZF) embryos have a small size (\SIrange{0.5}{1}{\mm}), demanding lower proton
energy and relatively small SOBP. Embryos correspond to functionally and morphologically organisms with the size of
organoids, well recognized as a good model for radiobiological evaluation of ionizing
radiation, particularly suitable for lesser penetrating
laser-accelerated
protons\cite{beyreutherFeasibilityProtonFLASH2019,roschFeasibilityStudyZebrafish2020,brunnerDosedependentChangesProton2020}.\\
In order to irradiate ZF embryos, the transport line was setup to produce a uniform
irradiation field over \SI{5}{\mm} diameter and \SI{600}{\um} depth in water. A specific
holder was designed to confine the embryos in close contact with the irradiation field
within the irradiation field surface while
preserving their survival conditions (see Methods).
The holder allows the irradiation of several embryos at the same time. Embryos were
irradiated at \SI{4}{hours} post-fertilization, when the embryo has a diameter of roughly
\SI{500}{\um}. Developed fish were fixated \SI{5}{days} post-irradiation, and the fish length measured
as an indicator of SP-LDP toxicity (\figref{fig:zf-length}).\\
Irradiated animals exhibited changes in morphology and in the length due to spine
curvature and developmental deterioration. 
As expected, after irradiation with SB-LDP, zebrafish embryos showed a significant
decrease in total length, with a length corresponding to \SI{75.2 +- 3.68}{\percent}
 of non-irradiated animals. As the total dose was deposited in a
single bunch, our irradiation conditions fall within the current definition of FLASH. It
is yet to verify if the much shorter duration and the much higher instantaneous dose rate
do have measurable effects in addition to those already demonstrated.
FLASH effect on zebrafish embryos was observed in similar conditions (\SI{8}{\gray} dose
	fraction at \SI{4}{\hour} post-fertilization, developmental evaluation at
\SI{5}{\day} post-irradiation) in Bourhis et al. 2019
paper\cite{bourhisClinicalTranslationFLASH2019}, where a \SI{25}{\percent} developmental
improvement is observed between FLASH and conventional irradiation conditions.
In our experiment, length decrease measured
for irradiated embryos was about \SI{25}{\percent} of the non irradiated for a slightly
higher dose (\SI{9,71 +- 1.37}{\gray}). This observation suggests that SP-LDP condition
could have triggered FLASH effect in the irradiated ZF embryos.

\section*{Discussion}
Due to their favorable ballistic properties, hadrons (protons and carbon ions)
represent a better alternative for the radiation therapy of solid tumors affecting
organ-at-risk. The Bragg peak dose deposition profile allows the dose to be better
concentrated within the target volume, reducing side effects to neighboring healthy
tissues. The development of laser plasma technology and its new paradigms for protons
acceleration (Wilks et al., 2001) raise new possibilities for studying the effects of dose
delivery modalities\cite{masoodCompactSolutionIon2014}.
However, due to the used particle (electrons or protons) acceleration technology
(laser-driven sources), the total deposited dose is delivered as a sequence of multiple
ultra-short separate fractions at ultra-high instantaneous dose rates; the dose per pulse
(hence the number of bunches required for a given dose) and the effective repetition rate
(hence the total irradiation time) depend on technical choices or limitations at the
facility\cite{binLaserdrivenNanosecondProton2012,doriaBiologicalEffectivenessLive2012,pommarelShapingControlLaseraccelerated2017,kraftDosedependentBiologicalDamage2010,raschkeUltrashortLaseracceleratedProton2016,yogoMeasurementRelativeBiological2011,zeilDosecontrolledIrradiationCancer2013}.
In the past we the radiobiological impact of the repetition rate of bunches and
we showed that, at constant dose, the temporal dose deposition modality of LDP
is a key parameter in determining cancer cells response\cite{bayartFastDoseFractionation2019}. \\
Since the first proof of concept of irradiating cells with
LDP\cite{kraftDosedependentBiologicalDamage2010,raschkeUltrashortLaseracceleratedProton2016} 
we have been able to explore the biological impact of
laser-accelerated protons of highly resistant glioblastoma cells, for which proton
therapy is one of the main indications. Cell
survival of U87-MG glioblastoma cell line showed, as expected, similar effectiveness of
single-pulse-LDP (SP-LDP), compared to fast fraction-LDP and conventional beams, to induce
cell killing (\figref{fig:u87-surv}). Indeed, we showed that this cell line was unsensitive to the
temporal parameter of dose deposition resulting from the absence of functional PARP1
protein which is probably responsible of its high radioresistance.
We also investigated the ability of laser-accelerated protons to generate
DNA double-strand breaks (DSBs); it is confirmed that LDP and conventional beams have
similar effectiveness, although a number of non-significant divergences were reported
(mainly arising from differences in experimental procedures and endpoints).\\
Flash effect and related healthy tissue sparing have been reported
to be linked to reduced production of reactive oxygen species
(ROS)\cite{wilsonRevisitingUltrahighDose2012,montay-gruelLongtermNeurocognitiveBenefits2019}.

In this study, the high charge generated by the high energy Pico2000 laser, using
the well-established TNSA acceleration technique, enabled irradiation conditions
compatible with the requirements of the FLASH protocol. Owing to the high energy/low
repetition rate configuration, the FLASH condition was reached within a single
\SI{1}{\ps} laser pulse; proton bunches duration is shorter than \SI{10}{\ns},
which implies a dose-rate exceeding \SI{e8}{\gray\per\s}.\\
The use of a quadrupole transport line, enabled to control the deposited dose within the
single proton pulse. In our setup, a moving scattering filter installed after the
second magnetic focusing element was used to adjust the diffusion center of proton
emission, thereby affecting the particle density at the irradiation plane. 
This technique enables dose escalation, even with the
ultra-fast nature of laser-driven acceleration mechanism. Additionally, the two focusing
elements naturally influence the particle spectrum by limiting the low energy end, thus
enhancing the dose contribution from the central portion. As a result, this effect
produced a more uniform SOBP through the \SI{1}{\mm} thick biological sample.\\
We demonstrated how the temperature of the TNSA proton emission could be used as a
parameter to fit the available diagnostics on the proton transport line. In fact the
otherwise non-optimal acceleration repeatability proved to loosely depend on temperature
and strongly on the amount of accelerated charge. The Analysis protocol we set up, not
only enabled a precise reconstruction of the dose within the target volume, but also to
extract beam parameters otherwise non accessible. This strategy will be of use for future
experiments where TNSA accelerated protons will be used for practical applications, most
importantly for laser-driven single-pulse FLASH irradiation condition. \\
Notwithstanding the challenges of retrieving irradiation conditions post-irradiation, our experiment
highlights a fundamental limitation of laser-driven single-pulse FLASH: it is not
possible to monitor, and therefore to control, the total deposited dose before the
irradiation is completed.\\

The experimental evidence we gathered under laser-driven FLASH conditions was used to
explore whether induction of oxidative stress-dependent DNA damage could differ between 
healthy and tumor cells under these irradiation modalities.  By comparing the ratios
between the two cell types, we observed that oxidative stress-dependent DNA damage
was equivalent in non-irradiated cells; however it was higher in U87-MG glioblastoma cells
than in healthy MRC5 fibroblasts (\figref{fig:oxy-mrc5-u87}). This result suggests that
applying SP-LDP may be able to generate a greater amount of oxidative stress-dependent DNA
damage in tumor model than in a healthy one. This hypothesis aligns with
recent data indicating that FLASH irradiation produces significantly more ROS than conventional
irradiation, which can be eliminated much more efficiently in normal tissues during steady-state metabolism than in tumors.

\bibliography{zotero} 

\section*{Acknowledgments}
AF acknowledges region Ile-de-France for the contract Sesame/IDRA, and ANR contract
FemtoDose. AF and CG acknowledge CNRS contract MITI/BioRapide.

\section*{Author contributions statement}
AF proposed the experiment in collaboration with MC and LR. TR, LT, JS and KP provided the
quadrupole system. AF, LR and EB performed the
experiment with support from MC, JdM, TR, AP. LdM and AP provided calibration of RCF. EB,
ILJ and CG prepared and analyzed biological samples. AF and EB wrote the paper with
contribution from LR, TR, CG, LdM, AP, KP and JS. 

\section*{Methods}

\subsection*{Dosimetry}
Spatial dose distribution is measured by means of radiochromic films (Ashland corp.) type
EBT3 and EBT3-XD. Films were calibrated on CPO medical accelerator with proton beam energy
degraded to \SI{20}{\MeV}. The calibration procedure is the same followed in
\cite{pommarelShapingControlLaseraccelerated2017,pommarelShapingControlLaserproduced2017}.

\subsection*{Monte Carlo simulation}
\begin{figure}[!ht]
		\begin{center}
			\setlength{\unitlength}{1cm}
			\begin{picture}(7,10)(0,0)
				\put(0,10.5){\bf (a)}
				\put(0.2,5.5){\includegraphics[width=7cm]{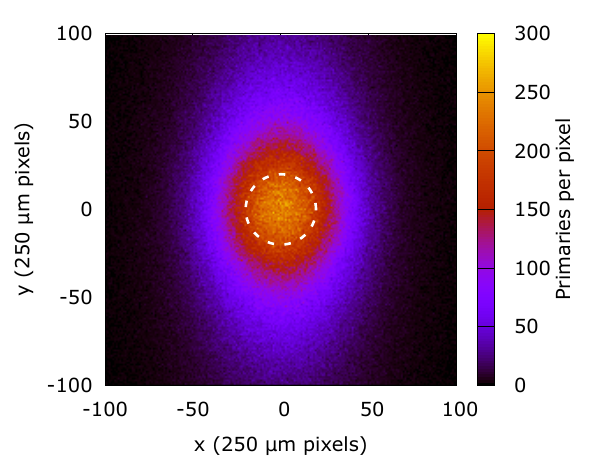}}

				\put(0,4.4){\bf (b)}
				\put(0,0){\includegraphics[width=7cm]{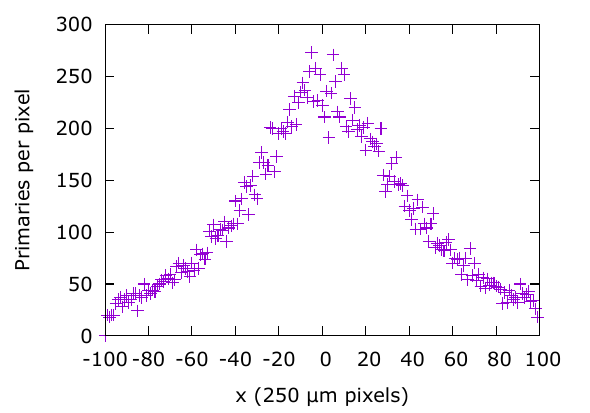}}
			\end{picture}
			\caption{Number of primaries in a simulation, map (a) and center cut-out (b)
			reaching the first RCF in a simulation with \SI{10}{\pico\coulomb} of
			primaries. There is a minimum of \SI{2e3}{primaries\per\mm\squared}
			(average: \SI{3.4e3}{primaries\per\mm\squared}), resulting in a
		total of \num{269512} primaries in the \SI{1}{\cm} diameter ROI.}
			\label{fig:particle-count}
		\end{center}
	\end{figure}

The proton beamline is simulated using the Monte Carlo toolkit Geant4, where quadrupoles
are represented
by measured field maps. The possibility of simulating magnetic optics with TNSA
sources and energies using Monte Carlo methods was studied in greater detail and validated in
Cavallone et al., 2019\cite{cavalloneShapingLaseracceleratedProton2019} by comparing it
with codes that account for space charge. \\
Monte Carlo simulations are run with a point proton source with an exponential spectrum as
in equation~\eqref{eq:spectrum}, where lower and higher cutoffs matching experimental
values. The lower cutoff is set to \SI{1}{\MeV}, which
is well below the minimum energy needed to reach the first radiochromic film in absence of
a scattering filter; while the maximum is set to \SI{19.5}{\MeV}, which corresponds to the
highest measured energy. Source has isotropic angular distribution up to $\theta_{max} =
\SI{100}{\milli\radian}$; source temperature E$_0$ and effective charge Q$_{0}^{*}$ are varied to
match dosimetry measurement. \\
The total proton charge on the test shot was estimated to be Q$_0 =
\SI{157}{\nano\coulomb}$, obtained from the exponential fitting algorithm described in
\cite{cavallone:tel-03085030}. The effective charge ratio (i.e. the total charge within
	the input acceptance angle of the first quadrupole) is calculated by numerical
	integration from fitted curves in \figref{fig:source-parameters} and found to
	be $\mathrm{Q}_{0}^{*} = 0.2\,\mathrm{Q}_{0} \simeq \SI{30}{\nC^{*}}$ which is used as an order of
	magnitude of the number of particles entering the transport line. This figure does
	not take in account the efficiency of the particle transport, nor the actual
	charge within the ROI on the biological sample. Given the numerical complexity of
	the transport line, the number of primaries in the Monte Carlo simulation is
	reduced to \SI{10}{\pico\coulomb^{*}} (roughly a factor \num{e-3} of the
	experimental charge) and then scaled back to \SI{1}{\nC^{*}}. A test simulation in
	the dose escalation configuration shows that, on the average, \SI{200}{primaries}
	are recorded per each \SI{250}{\um} pixel at the first RCF
	(\figref{fig:particle-count}).\\

EBT radiochromic films are represented in the simulation by
\SI{275}{\um} of composite material with correct relative atomic abundances and density.
For a setup where multiple RCF are present, separate simulations are run where each RCF
material is switched to water, in order to account for the equivalent deposited
dose-to-water. In fact, the RCF calibration is obtained as a water
equivalent (density $\rho_{water}=\SI{1}{\g\per\cm^3}$) while the actual average density of the
radiochromic foil is $\rho_{rcf} = \eta_{PE}\rho_{PE} + \eta_{Sens}\rho_{Sens} =
\SI{1.375}{\g\per\cm^3}$, where $\eta_{PE}=0.9$ and $\eta_{Sens}=0.1$ are respectively the
polyester and the sensitive layer volume fractions in a radiochromic EBT3 type foil. This
correction is needed in order to correctly simulate the dose deposition of low kinetic energy
protons in the sensitive regions of dosimeters and biological sample. The observed
correction due to the change water/polyesther can be as high as \SI{10}{\percent}.\\

\subsection*{Radiobiology procedures}

\subsubsection*{Cell culture}
The human glioblastoma cell line U87-MG was cultured in Dulbecco's modified Eagle's
minimum medium with Glutamax (Thermo Fisher Scientific). Cells were grown as monolayers,
supplemented with \SI{10}{\percent} fetal calf serum (PAA) and \SI{1}{\percent} penicillin
and streptomycin (ThermoFisher Scientific) in plastic tissue culture disposable flasks
(TPP) at \SI{37}{\celsius} in a humidified atmosphere of \SI{5}{\percent} $\mathrm{CO_2}$
in air.

\subsubsection*{Cell survival assay}
The cell containers used for irradiation were lumox\textsuperscript{\textregistered} dish 35 (SARSTEDT)
exhibiting a \SI{25}{\um} thick lumox\textsuperscript{\textregistered} bottom face. Depending on beam
transverse profile and position observed on radiochromic films (ref) a \SI{1}{\cm}
circular area was delimited on the internal face of the lumox\textsuperscript{\textregistered} membrane,
where $3 \times 10^4$ cells were seeded. Cells were let grow overnight in \SI{200}{\ul}  of
medium. To generate dose-response survival curves U87-MG cell lines were subjected to
doses varying from \SIrange{2.5}{10.8}{\gray} with single bunch of laser driven protons.
Survival curves resulting from irradiation with CAP or fast-fractionation LDP were
retrieved from \cite{bayartFastDoseFractionation2019}. After exposure to ionizing radiations,
cells were incubated for \SI{3}{\hour} in standard conditions. Cells were harvested with
Accutase (Merck), dispatched into \num{3} different wells of \num{12}-well plates (TPP) in
\SI{2.5}{\ml} of medium and grown for five generations corresponding to \SI{6}{\day} for
U87-MG cell line. Cells were harvested with Accutase which was then inactivated using an
equal volume of $1X$ PBS (ThermoFisher Scientific) supplemented with \SI{10}{\percent}
fetal calf serum. The final volume was adjusted to \SI{1}{\ml} with $1X$ PBS and
\SI{200}{\ul} of each well were dispatched into a non-sterile U-bottom \num{96}-well plate
(TPP). In each well, \SI{2}{\ul} of a propidium iodide solution (Sigma, 100 \si{\ug/\ml}
in 1X PBS) were added just before flow cytometry counting. Cell acquisition and data
analysis were performed using Guava\textsuperscript{\textregistered} and GuavaSoft (Merck) and then GraphPad Prism software.

\subsubsection*{Oxidative stress-dependent DNA damage analysis}
To perform oxidative stress-dependent DNA damage analysis, \num{6e4} cells from the MRC5
healthy cell line and U87-MG glioblastoma cell line were seeded in the same cell
container, as describe above. Cells were irradiated at a target dose of \SI{2.5}{\gray},
harvested one hour post-irradiation and immediately frozen in liquid nitrogen before being
stored at \SI{-80}{\celsius}. Oxidative stress-dependent DNA damage (8-hydroxyguanosine, 8-OHdG)
measurement was performed using DNA damage ELISA kit (Enzo) following manufacturer
protocol. Absorbance at \SI{450}{\nm} was measured using an EnSpire Plate reader
(PerkingElmer), and the data were analyzed using GraphPad Prism software.

\subsubsection*{Zebrafish holder}
\begin{figure}[!ht]
	\begin{center}
		\setlength{\unitlength}{1cm}
		\begin{picture}(15, 5.2)(0,0)
			\put(0,0){\includegraphics[width=9cm]{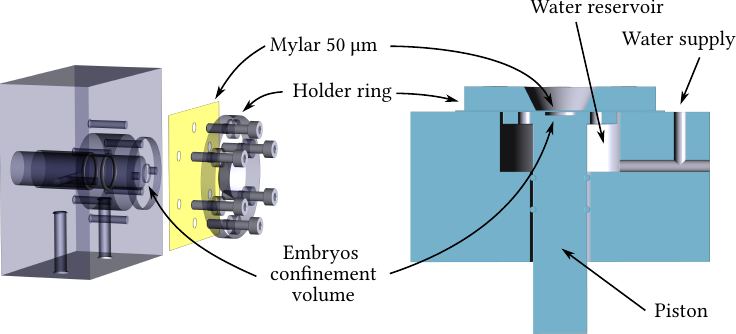}}
			\put(0,0){\bf (a)}
			\put(5.7,0){\bf (b)}
		\end{picture}
		\caption{Exploded view (a) and sagittal cut-out (b) of the holder developed
		for the irradiation of Zebrafish embryos. The piston run through the
		water reservoir is visible, for switching between storage and irradiation
		configuration. }
		\label{fig:zf-mount}
	\end{center}
\end{figure}

Owing to the limited available kinetic energy in the proton beam, a specific holder
(\figref{fig:zf-mount})
for
zebrafish embryos was designed and fabricated. The embryos are kept in a water
reservoir before irradiation; the holder is kept horizontal (as in
\figref{fig:zf-mount}-b) and the embryos are let sink
in a \SI{600}{\um} deep notch on the top of the piston. At the moment of irradiation the
piston is pushed upwards towards the Mylar water/air separation. In this condition embryos
are confined in the inner cylinder, which makes possible to put the holder in vertical
position (as in \figref{fig:zf-mount}-a) for installation on the irradiation beam-line.

\subsubsection*{Zebrafish embryos experiments}
For in vivo studies, wild-type (WT) zebrafish were bred in LOB fish facility (École
Polytechnique). All in vivo experiments on zebrafish were performed on embryos below
\num{5} days post-irradiation. Zebrafish embryos were obtained by natural spawning of WT fish.
Fertilized WT zebrafish eggs were incubated at \SI{28}{\celsius} until \SI{5}{days}
post-irradiation. Irradiation was
performed \SI{4}{hours\,post-fertilization} at a target dose of \SI{10}{\gray} SP-LDP and
FLASH (Oriatron, Institut Curie, as control) in zebrafish holder. For FLASH irradation,
zebrafish holder was placed in a water tank. Embryos were fixed \SI{5}{days}
post-irradiation with a solution of paraformaldehyde
(\SI{4}{\percent} final concentration for one hour) before microscopic analysis (Evos XL
Core Cell Imaging System; Thermo Fisher Scientific). Fish length was measured using ImageJ
1.X. software.

\end{document}